\documentclass[twocolumn,showpacs,prl,superscriptaddress,a4paper,floatfix]{revtex4}

\usepackage{graphicx}
\usepackage{amsmath}
\usepackage{graphicx}
\usepackage{dcolumn}
\usepackage{bm}

\begin{document}

\title{Decoherence by a spin thermal bath: Role of the spin-spin interactions}
\author{Shengjun Yuan}
\affiliation{Department of Applied Physics, Zernike Institute for Advanced Materials,
University of Groningen, Nijenborgh 4, NL-9747 AG Groningen, The Netherlands}
\author{Mikhail I. Katsnelson}
\affiliation{Institute of Molecules and Materials, Radboud University of Nijmegen, 6525
ED Nijmegen, The Netherlands}
\author{Hans De Raedt}
\affiliation{Department of Applied Physics, Zernike Institute for Advanced Materials,
University of Groningen, Nijenborgh 4, NL-9747 AG Groningen, The Netherlands}
\date{\today }

\begin{abstract}
We study the decoherence of two coupled spins that interact with a
spin-bath environment. It is shown that the connectivity
and the coupling strength between the spins in the environment are of
crucial importance for the decoherence of the central system.
Changing the connectivity or coupling strenghts changes
the decoherence of the central system from Gaussian to exponential decay law.
\end{abstract}

\maketitle





It is commonly accepted that decoherence by nuclear spins is the main
obstacle for realization of quantum computations in magnetic systems; see,
e.g., discussions of specific silicon \cite{kane} and carbon \cite{loss}
based quantum computers. Therefore, understanding the decoherence in quantum
spin systems is a subject of numerous works (for reviews, see Refs~\cite{stamp,ZhangW2007}).
The issue seems to be very complicated and despite many
efforts, even some basic questions about character of the decoherence
process are unsolved yet.
Due to the interactions with and between the spin of the bath,
an analytical treatment can be carried out in very exceptional cases,
even if the central systems contains one spin only.
Recent work suggests that the internal dynamics
of the environment can be crucial to the decoherence of the central
system~\cite{Dawson2005,Rossini2007,Tessieri2003,Camalet2007,YuanXZ2005,
YuanXZ2007, Wezel2005,Bhaktavatsala2007,ourPRE,Relano2007,ZhangW2006,
JETPLett,Yuan2007}. In this Letter, we present results of extensive
simulation work of a two-spin system interacting with a spin-bath
environment and show that the decoherence of the two-psin system
can exhibit different behavior, depending on the characteristics
of the coupling with the environment and of the internal dynamics
of the latter. We also provide a simple physical picture to understand
this behavior.

In general, the behavior of an open quantum system crucially depends on the ratio of
typical energy differences of the central system $\delta E_c$ and the energy
$E_{ce}$ which characterizes the interaction of the central system with the
environment. The case $\delta E_c \ll E_{ce}$ has been studied extensively
in relation to the ``Schr\"{o}dinger cat'' problem and the physics is quite
clear~ \cite{zeh,zurek}: As a result of time evolution, the central system
passes to one of the ``pointer states''~\cite{zurek} which, in this case,
are the eigenstates of the interaction Hamiltonian $H_{ce}$. The opposite case,
$\delta E_c \gg E_{ce}$ is less well understood. There is a conjecture that
in this case the pointer states should be eigenstates of the Hamiltonian $H_c$ of
the central system but this has been proven for a very simple model only~\cite{paz}.
On the other hand, this case is of primary interest if, say, the
central system consists of electron spins whereas the environment are
nuclear spins, for instance if one considers the possibility of quantum computation
using molecular magnets~\cite{m1,m2}.

We consider a generic quantum spin model described by the Hamiltonian
$H=H_{c}+H_{ce}+H_{e}$ where
$H_{c} =-J\mathbf{S}_{1}\cdot \mathbf{S}_{2}$ is the Hamiltonian
of the central system and
the Hamiltonians of the environment and the
interaction of the central system with the environment are given by
\begin{eqnarray}
H_{e}&=&-\sum_{i=1}^{N-1}
\sum_{j=i+1}^{N}\sum_{\alpha }\Omega _{i,j}^{(\alpha )}I_{i}^{\alpha
}I_{j}^{\alpha },
\nonumber \\
H_{ce}& =&-\sum_{i=1}^{2}\sum_{j=1}^{N}\sum_{\alpha }\Delta _{i,j}^{(\alpha
)}S_{i}^{\alpha }I_{j}^{\alpha },
\label{HAM}
\end{eqnarray}
respectively. The exchange integrals $J$ and $\Omega _{i,j}^{(\alpha )}$
determine the strength of the interaction between spins $\mathbf{S}%
_{n}=(S_{n}^{x},S_{n}^{y},S_{n}^{z})$ of the central system, and the
spins $\mathbf{I}_{n}=(I_{n}^{x},I_{n}^{y},I_{n}^{z})$ in the environment, respectively.
The exchange integrals $\Delta _{i,j}^{(\alpha )}$
control the interaction of the central system with its environment.
In Eq.~(\ref{HAM}), the sum over $\alpha $ runs over the $x$, $y$ and $z$
components of spin-$1/2$ operators $\mathbf{S}$ and $\mathbf{I}$. In the
sequel, we will use the term \textquotedblleft
Heisenberg-like\textquotedblright\ $H_{e}$ ($H_{ce}$) to indicate that each $%
\Omega _{i,j}^{(\alpha )}$ ($\Delta _{i,j}^{(\alpha )}$) is a uniform random
number in the range $[-\Omega |,\Omega ]$ ($[-\Delta ,\Delta ]$), $\Omega $ and
$\Delta $ being free parameters. In earlier work~\cite{JETPLett,Yuan2007},
we found that a Heisenberg-like $H_{e}$ can induce close to perfect decoherence of the
central system and therefore, we will focus on this case only.

The bath is further characterized by the number of environment spins $K$
with which a spin in the environment interacts.
If $K=0$, each spin in the environment interacts with the central system only.
$K=2$, $K=4$ or $K=6$ correspond to environments in which
the spins are placed on a ring, square or triangular lattice, respectively and
interact with nearest-neighbors only.
If $K=N-1$, each spin in the environment interacts with all the other spins in the environment and, to
give this case a name, we will refer to this case as \textquotedblleft spin
glass\textquotedblright .

If the Hamiltonian of the central system $H_{c}$ is a perturbation relative
to the interaction Hamiltonian $H_{ce}$, the pointer states are eigenstates
of $H_{ce}$~\cite{zurek}. In the opposite case, that is the regime $|\Delta
|\ll |J|$ that we explore in this Letter, the pointer states are conjectured to
be eigenstates of $H_{c}$~\cite{paz}. The latter are given by $|1\rangle
\equiv |T_{1}\rangle =\left\vert \uparrow \uparrow \right\rangle $, $%
|2\rangle \equiv |S\rangle =(\left\vert \uparrow \downarrow \right\rangle
-\left\vert \downarrow \uparrow \right\rangle )/\sqrt{2}$, $|3\rangle \equiv
|T_{0}\rangle =(\left\vert \uparrow \downarrow \right\rangle +\left\vert
\downarrow \uparrow \right\rangle )/\sqrt{2}$, and $|4\rangle \equiv
|T_{-1}\rangle =\left\vert \downarrow \downarrow \right\rangle $, satisfying
$H_{c}|S\rangle =(3J/4)|S\rangle $ and $H_{c}|T_{i}\rangle
=(-J/4)|T_{i}\rangle $ for $i=-1,0,1$.

The simulation procedure is as follows. We generate a random superposition $%
\left\vert \phi \right\rangle $ of all the basis states of the environment.
This state corresponds to the equilibrium density matrix of the environment
at infinite temperature. The spin-up -- spin-down state ($\left\vert
\uparrow \downarrow \right\rangle $) is taken as the initial state of the
central system. Thus, the initial state of the whole system reads $%
\left\vert \Psi (t=0)\right\rangle \rangle =\left\vert \uparrow \downarrow
\right\rangle \left\vert \phi \right\rangle $ and is a product state of the
state of the central system and the random state of the environment which,
in general is a (very complicated) linear combination of the $2^{N}$ basis
states of the environment. In our simulations we take $N=16$ which, from
earlier work~\cite{JETPLett,Yuan2007}, is sufficiently large for the
environment to behave as a \textquotedblleft large\textquotedblright\
system. For a given, fixed set of model parameters, the time evolution of
the whole system is obtained by solving the time-dependent Schr\"{o}dinger
equation 
for the many-body wave function $|\Psi (t)\rangle $, describing the central
system plus the environment~\cite{method}. It conserves the energy of the whole system to
machine precision. We monitor the effects of the decoherence by computing the
the matrix elements of the reduced density matrix $\rho\left( t\right) $
of the central system.
As explained earlier, in the regime
of interest $|\Delta |\ll |J|$, the pointer states are expected to be the
eigenstates of the central systems.
Hence we compute the matrix elements of
the density matrix in the basis of eigenvectors of the central system.
We also compute the time dependence of quadratic entropy $S_{c}\left( t\right)
=1-Tr\rho ^{2}\left( t\right) $ and the Loschmidt echo $L\left( t\right)
=Tr\left( \rho \left( t\right) \rho _{0}\left( t\right) \right) $~\cite{Cucchietti2003},
where $\rho _{0}\left( t\right) $ is the density matrix for $H_{ce}=0$.

\begin{figure}[t]
\begin{center}
\includegraphics[width=9cm]{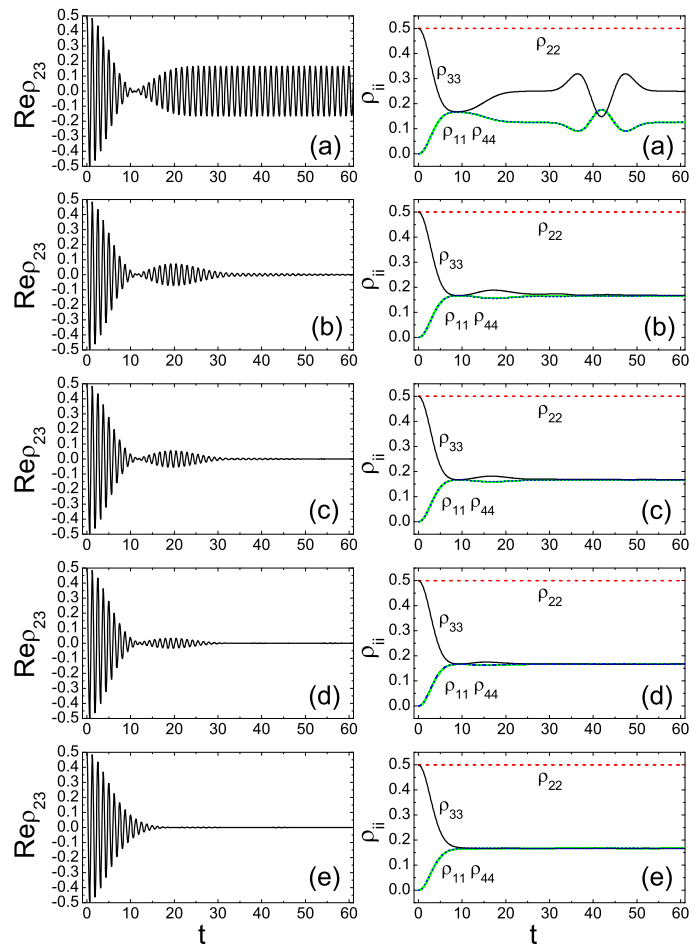}
\end{center}
\caption{(Color online) The time evolution of the real part of the
off-diagonal element $\protect\rho _{23}$ (left panel) and the diagonal
elements $\protect\rho _{11},\ldots ,\protect\rho _{44}$ (right panel) of the
reduced density matrix of a central system (with $J=-5$), coupled
via an isotropic Heisenberg interaction $H_{ce}$ ($\Delta =-0.075$ )
to a Heisenberg-like environment $H_{e}$ ($\Omega =0.15$) with different connectivity:
(a) $K=0$; (b) $K=2$; (c) $K=4$; (d) $K=6$; (e) $K=N-1$. }
\label{fig1}
\end{figure}

\begin{figure}[t]
\begin{center}
\includegraphics[width=8.5cm]{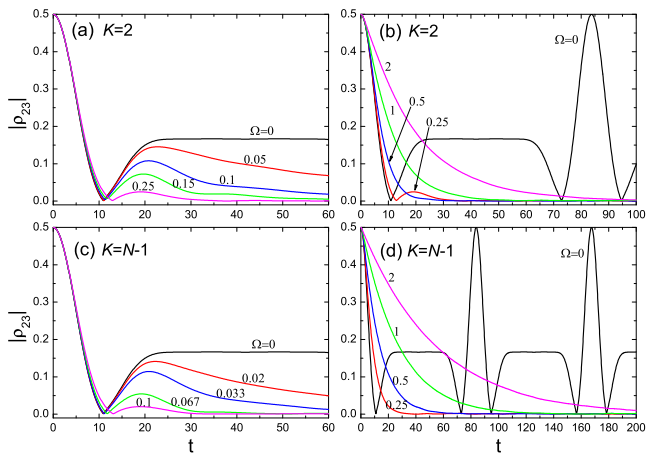}
\end{center}
\caption{(Color online) The time evolution of the off-diagonal element $%
\protect\rho _{23}$ of the reduced density matrix of a central system (with $%
J=-5$), interacting with a Heisenberg-like environment $H_{e}$
via an isotropic Heisenberg Hamiltonian $H_{ce}$ (with $%
\Delta =-0.075$ ) for the same geometric structures in the environment:
(a,b) $K=2$ and (c,d) $K=N-1$. The number next to each curve is the
corresponding value of $\Omega $.}
\label{fig2}
\end{figure}

In Fig.~\ref{fig1} and Fig.~\ref{fig2}, we show the time evolution of the
elements of the reduced density matrix $\rho \left( t\right) $ for
different connectivity $K$ and $\Omega$, for the case that $H_{ce}$
is an isotropic Heisenberg model.

If $\left\vert \Delta \right\vert \gg K\Omega $, in agreement with
earlier work~\cite{ourPRL,Melikidze2004}, we find that in the absence of
interactions between the environment spins ($K\Omega =0$) and after the
initial decay, the central system exhibits long-time oscillations (see Fig.~\ref{fig1}(a)(left)).
In this case and in the limit of a large environment, we have~\cite{Melikidze2004}
\begin{equation}
\hbox{Re }\rho _{23}\left( t\right) =\left[ \frac{1}{6}+\frac{1-bt^{2}}{3}%
e^{-ct^{2}}\right] \cos \omega t,  \label{melik}
\end{equation}%
where $b=N\Delta ^{2}/4$, $c=b/2$ and $\omega =J-\Delta $. Equation~(\ref{melik})
clearly shows the two-step process, that is, after the initial
Gaussian decay of the amplitude of the oscillations, the oscillations revive
and their amplitude levels of~\cite{Melikidze2004}.
Due to conservation laws, this behavior does not change if the environment
consists of an isotropic Heisenberg system ($\Omega _{i,j}^{(\alpha)}\equiv \Omega $
for all $\alpha $, $i$ and $j$), independent of $K$.
If, as in Ref.~\cite{ourPRL}, we take
$\Delta _{i,j}^{(x)}=\Delta_{i,j}^{(y)}=\Delta _{i,j}^{(z)}\in \left[ 0,\Delta \right] $
random instead of the identical, the amplitude of the long-living oscillations is no longer
constant but decays very slowly~\cite{ourPRL}.

If $\left\vert \Delta \right\vert \approx K\Omega $, the presence of
Heisenberg-like interactions between the spins of
the environment has little effect on the initial Gaussian decay of the
central system, but it leads to a reduction and to a decay of the amplitude of
the long-living oscillations. The larger $K$ (see
Fig.~\ref{fig1}(b-e)(left)) or $\Omega $ ({see Fig.~\ref{fig2}(a,c)}),
the faster the decay is.
Note that for the sake of clarity, we have suppressed the
fast oscillations by plotting instead of the real part, the absolute value
of the matrix elements.

If $\left\vert \Delta \right\vert \ll K\Omega $, keeping $K$ fixed
and increasing $\Omega $ smoothly changes the initial decay from
Gaussian (fast) to exponential (slow), and the long-living oscillations
are completely suppressed ({see Fig.~\ref{fig2}(b,d)}).
For large $\Omega $, the simulation data fits very well to
\begin{equation}
\left\vert \rho _{23}\left( t\right) \right\vert =\frac{1}{2}e^{-A_{K}\left(
\Omega \right) t},  \label{p23_exp}
\end{equation}%
with $A_{K}\left( \Omega \right) \approx \Omega\widetilde{A}_{K}  $,
$\widetilde{A}_{2}=9.13$ and $\widetilde{A}_{N-1}=26.73$.

\begin{figure}[t]
\begin{center}
\includegraphics[width=8.5cm]{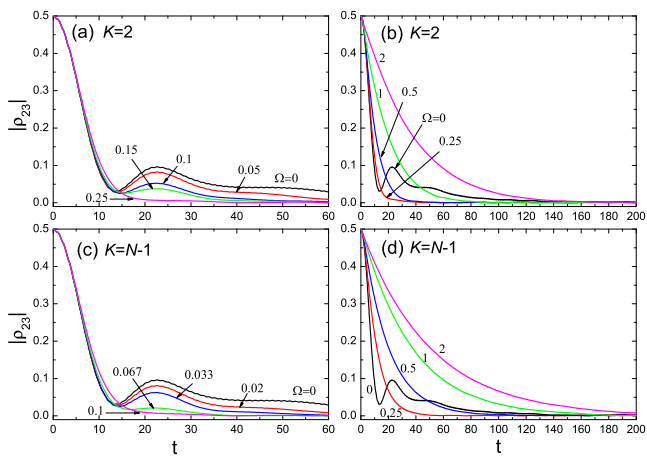}
\end{center}
\caption{(Color online) Same as Fig.~\ref{fig2}
except that $H_{ce}$ is Heisenberg-like and  $\Delta =0.15$.
}
\label{fig3}
\end{figure}

Physically, the observed behavior can be understood as follows.
If $\left\vert \Delta \right\vert \approx K\Omega $,
a bath spin is affected by roughly the same amount
by the motion of both the other bath spins and by the two central spins.
Therefore, each bath spin has enough freedom to follow the
original dynamics, much as if there were no coupling between bath spins.
This explains why the initial Gaussian decay is insensitive to
the values of $K$ or $\Omega $.
After the initial decay, the whole system is expected to reach an stationary state,
but because of the presence of Heisenberg-like interactions between the bath spins, a new
stationary state of the bath is established, suppressing the long-living oscillations.

For increasing $K$, the distance between two bath spins,
defined as the minimum number of bonds connecting the two spins, becomes smaller.
For instance, for $K=2$, this distance is $\left( N-2\right) /2$, and for $K=N-1$, it is zero.
Therefore, for fixed $\Omega $ and increasing $K$
the fluctuations in the spin bath can propagate faster and the
evolution to the stationary state will be faster.
Similarly, for fixed $K$,
increasing the coupling strength between the bath spins will speed up the
dynamics of the bath, that is, the larger $\Omega $ the faster will be the evolution to
the stationary state.

In the opposite case $\left\vert \Delta \right\vert\ll K\Omega $,
$H_{ce}$ is a small perturbation relative to $H_{e}$
and the coupling between bath spins is the dominant factor
in determining the dynamics of the bath spins.
Therefore, by increasing $K$ or $\Omega $, the bath spin will have less
freedom to follow the dynamics induced by the coupling to the two central spins,
the influence of the bath on the central system will decrease, and the
(exponential) decay will become slower.

According to the general picture of decoherence~\cite{zurek}, for an
environment with nontrivial internal dynamics that initially is in a random
superposition of all its eigenstates, we expect that the central system will
evolve to a stable mixture of its eigenstates. In other words, the
decoherence will cause all the off-diagonal elements of the reduced density
matrix to vanish with time. In the case of an isotropic Heisenberg coupling
between the central system and the environment, $H_{c}$ commutes
with the Hamiltonian $H$, hence the energy of the central system is a
conserved quantity.
Therefore, the weight of the singlet $\left\vert {S}\right\rangle $
in the mixed state should be a constant ($1/2$), and
the weights of the degenerate eigenstates $|T_{0}\rangle $, $|T_{-1}\rangle $
and $|T_{1}\rangle $ are expected to become the same (${1/6}$).
As shown in Fig.~\ref{fig1}(b-e)(right), our simulations confirm that this picture is
correct in all respects.

It is important to note that although in the foregoing discussion
we have compared $K\Omega $ to $\left\vert \Delta \right\vert $,
this does not imply that $K\Omega $ can be used to fully characterize
the decoherence process.
In order to clarify the role of $K$ and $\Omega $, we change the coupling
between the central system and the bath from Heisenberg to Heisenberg-like.
From a comparison of the data in Fig.~\ref{fig2} and Fig.~\ref{fig3},
it is clear that the roles of $K$ and $\Omega $ are the same in both cases,
no matter whether the central-bath coupling is isotropic or anisotropic.
However, there are some differences in the decoherence process.

If $\left\vert \Delta\right\vert \gg K\Omega $,
in the presence of anisotropic interactions between the central system
and the environment spins, even in the absence of
interactions between the bath spins,
the second step of the oscillations
decays and finally disappear as $K$ increases.
This is because the anisotropic
interactions break the rotational symmetry of the coupling between central system and environment
which is required for the long-living oscillations to persist.

If $\left\vert\Delta \right\vert \ll K\Omega $, $\left\vert \rho _{23}\left(t\right) \right\vert $
can still be described by Eq.~(\ref{p23_exp}), but now $A_{K}\left( \Omega \right) $
is no longer a linear function of $\Omega $.
For an anisotropic $H_{ce}$, the energy of the
central system is no longer a conserved quantity.
Therefore there will be energy transfer between the central system and the
environment and the weight of each pointer state (eigenstate) in the final
stable mixture need not be the same for all $K$ or $\Omega $.

\begin{figure}[t]
\begin{center}
\includegraphics[width=8.5cm]{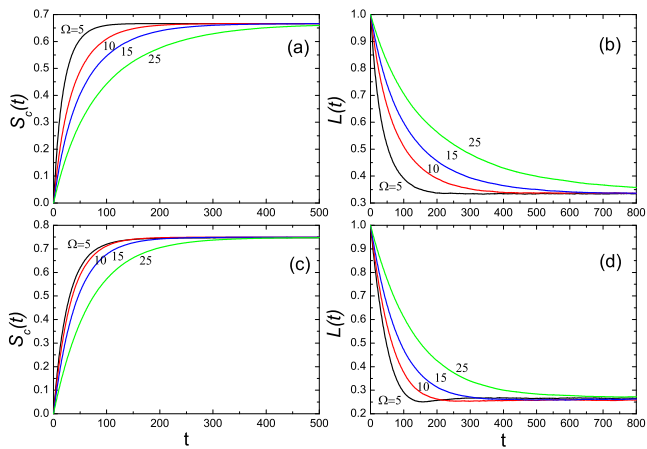}
\end{center}
\caption{(Color online) The time evolution of the the entropy $S_{c}\left(
t\right) $ and Loschmidt echo $L\left( t\right) $ of a central system (with $%
J=-5$), interacting with a Heisenberg-like environment $H_{e}$ (with
different $\Omega $) via a Heisenberg (a,b, $\Delta =-0.075$) or
Heisenberg-like (c,d, $\Delta =0.15$) Hamiltonian $H_{ce}$ for the case $K=2$.
The number next to each curve is the corresponding value of $\Omega $.}
\label{fig4}
\end{figure}

For a change, we illustrate this point by considering
the quadratic entropy $S_{c}\left( t\right) $ and Loschmidt echo $L\left( t\right) $.
We expect that these quantities will also dependent of the symmetry
of the coupling between central system and the spin bath.
In Fig.~\ref{fig4}, we present results for large $\Omega $ and $K=2$,
confirming this expectation.
For isotropic (Heisenberg) $H_{ce}$ and perfect decoherence
(zero off-diagonal terms in the reduced density matrix)
we expect that
$\max_t S_{c}(t)=1-[\left( 1/2\right) ^{2}+3\times \left( {1/6}\right) ^{2}]=2/3$,
in concert with the data of Fig.~\ref{fig4}(a)).
For Heisenberg-like $H_{ce}$, 
$\max_t S_{c }(t)$ will depend on the coupling strengths
and as shown in Fig.~\ref{fig4}(c), we find that $\max_t S_{c }(t) =$ $1-4\times \left( {1/4}\right) ^{2}=3/4$,
corresponding to the case that all the diagonal elements in the
reduced density matrix are the same ($1/4$) and all other elements are zero.

In conclusion, we have shown how a pure quantum state of the central
spin system evolves into a mixed state, and that if the
interaction between the central system and environment is much smaller than
the coupling between the spins in the central system, the pointer states are
the eigenstates of the central system. Both these observations are in
concert with the general picture of decoherence~\cite{zurek}.
Furthermore, we have demonstrated that, in the case that the environment is
a spin system, the details of this spin system are important for the
decoherence of the central system. In particular, we have shown that
changing the internal dynamics of the environment (geometric structure or
exchange couplings) may change the decoherence of the central spin system from Gaussian to exponential decay.

\section*{Acknowledgement}

M.I.K. acknowledges a support by the Stichting Fundamenteel Onderzoek der
Materie (FOM).


\begin{thebibliography}{99}
\bibitem{kane} B.E. Kane, Nature \textbf{393}, 133 (1998).

\bibitem{loss} B. Trauzettel \textit{et al}., Nature Phys. \textbf{3}, 192 (2007).

\bibitem{stamp} N.V. Prokof'ev and P.C.E. Stamp, Rep. Prog. Phys. \textbf{%
63}, 669 (2000).

\bibitem{ZhangW2007} W. Zhang \textit{et al}., J. Phys.: Cond. Matter
\textbf{19}, 083202 (2007).

\bibitem{Dawson2005} C.M. Dawson \textit{et al}., Phys. Rev A 71, 052321
(2005).

\bibitem{Rossini2007} D. Rossini \textit{et al}., Phys. Rev. A \textbf{75},
032333 (2007).

\bibitem{Tessieri2003} L. Tessieri and J. Wilkie, J. Phys. A: Math. Gen.
\textbf{36}, 12305 (2003).

\bibitem{Camalet2007} S. Camalet and R. Chitra, Phys. Rev. B \textbf{75},
094434 (2007).

\bibitem{YuanXZ2007} X.Z. Yuan, H-S Goan and K.D. Zhu, Phys. Rev. B
\textbf{75}, 045331 (2007).

\bibitem{YuanXZ2005} X.Z. Yuan and K.D. Zhu, Europhys. Lett. \textbf{69},
868 (2005).

\bibitem{Wezel2005} J. van Wezel, J. van den Brink and J. Zaanen, Phys. Rev.
Lett. \textbf{94}, 230401 (2005).

\bibitem{Bhaktavatsala2007} D.D. Bhaktavatsala Rao \textit{et al}., Phys.
Rev. A \textbf{75}, 052338 (2007).

\bibitem{ourPRE} J. Lages \textit{et al}., Phys. Rev. E \textbf{72}, 026225
(2005).

\bibitem{Relano2007} A. Relano, J. Dukelsky and R.A. Molina, arXiv:0709.1383.

\bibitem{ZhangW2006} W. Zhang \textit{et al}., Phys. Rev. B \textbf{74},
205313 (2006).

\bibitem{JETPLett} S. Yuan, M.I. Katsnelson and H. De Raedt, JETP Lett.
\textbf{84}, 99 (2006).

\bibitem{Yuan2007} S. Yuan, M.I. Katsnelson and H. De Raedt, Phys. Rev. A
\textbf{75}, 052109 (2007).


\bibitem{zeh} D. Giulini \textit{et al}., \textit{Decoherence and the
Appearance of a Classical World in Quantum Theory} (Springer, Berlin, 1996).

\bibitem{zurek} W.H. Zurek, Rev. Mod. Phys. \textbf{75}, 715 (2003).

\bibitem{paz} J.P. Paz and W.H. Zurek, Phys. Rev. Lett. \textbf{82}, 5181
(1999).

\bibitem{m1} V.V. Dobrovitski , M.I. Katsnelson and B.N. Harmon, Phys.
Rev. Lett. \textbf{84}, 3458 (2000).

\bibitem{m2} M.N. Leuenberger and D. Loss, Nature \textbf{410}, 789 (2001).

\bibitem{method} V.V. Dobrovitski and H.A. De Raedt, Phys. Rev. E \textbf{%
67}, 056702 (2003).

\bibitem{Cucchietti2003} F.M. Cucchietti \textit{et al}.. Phys. Rev. Lett.
\textbf{91}, 210403 (2003).

\bibitem{ourPRL} V.V. Dobrovitski \textit{et al}., Phys. Rev. Lett. \textbf{%
90}, 210401 (2003).

\bibitem{Melikidze2004} A. Melikidze \textit{et al}., Phys. Rev. B \textbf{70%
}, 014435 (2004).

\end{thebibliography}
\end{document}